\documentclass[preprintnumbers, twocolumn,amsmath, amssymb, superscriptaddress, pra]{revtex4}
\usepackage{graphicx}
\usepackage{subfigure}
\usepackage{dcolumn}% Align table columns on decimal point
\usepackage{bm}% bold math
\usepackage{epsf}
\usepackage{amssymb}
\DeclareGraphicsExtensions{.pdf}
\begin{document}

%\normalsize{D.S. Simon$^1$, A.V. Sergienko$^{1,2}$}\\
%\small{1. Dept. of Electrical and Computer Engineering, Boston University, 8 Saint Mary's Street, Boston, MA 02215\\
%2. Dept. of Physics, Boston University, 590 Commonwealth Ave., Boston, MA 02215}
%\author{D.S. Simon}
%\affiliation{Dept. of Electrical and Computer Engineering, Boston
%University, 8 Saint Mary's St., Boston, MA 02215}
%\author{A.V. Sergienko}
%\affiliation{Dept. of Electrical and Computer Engineering, Boston
%University, 8 Saint Mary's St., Boston, MA 02215}
%\affiliation{Dept. of Physics, Boston University, 590 Commonwealth
%Ave., Boston, MA 02215}

\author{David S. Simon}
\email[e-mail: ]{simond@bu.edu} \affiliation{Dept. of Physics and Astronomy, Stonehill College, 320 Washington Street, Easton, MA 02357} \affiliation{Dept. of
Electrical and Computer Engineering \& Photonics Center, Boston University, 8 Saint Mary's St., Boston, MA 02215, USA}
\author{Shuto Osawa}
\email[e-mail: ]{sosawa@bu.edu} \affiliation{Dept. of Electrical and Computer Engineering \& Photonics Center, Boston University, 8 Saint Mary's St., Boston, MA
02215, USA}
\author{Alexander V. Sergienko}
\email[e-mail: ]{alexserg@bu.edu} \affiliation{Dept. of Electrical and Computer Engineering \& Photonics Center, Boston University, 8 Saint Mary's St., Boston,
MA 02215, USA} \affiliation{Dept. of Physics, Boston University, 590 Commonwealth Ave., Boston, MA 02215, USA}

\begin{abstract}
We demonstrate a previously unknown two-photon effect in
a discrete-time quantum walk. Two identical bosons with no mutual interactions nonetheless can remain clustered together as they walk on a lattice of
\emph{directionally-reversible} optical four-ports acting as Grover coins; both photons move in the \emph{same} direction at each step due to a two-photon quantum interference phenomenon
reminiscent of the Hong-Ou-Mandel effect. The clustered two-photon amplitude splits into two
localized parts, one oscillating near the initial point, and the other moving ballistically without spatial
spread, in soliton-like fashion. But the two photons are always clustered in the \emph{same} part of the superposition, leading to potential applications for transport of entanglement and opportunities for novel two-photon interferometry experiments.
\end{abstract}

\title{Quantum-Clustered Two-Photon Walks}

%\pacs{42.50.St,42.15.Fr,42.50.Dv,42.30.Kq}

\maketitle
%
%\begin{center}{\large\bf Quantum-Clustered Two-Photon Walks and Non-Stationary Two-Photon Cat States: A Quantum Walk Analog of the HOM Effect}\\ \vskip 4pt
%\today\end{center}

\section{Introduction} 
\begin{figure}[b!]
\begin{center}
\subfigure[]{
\includegraphics[height=1.3in]{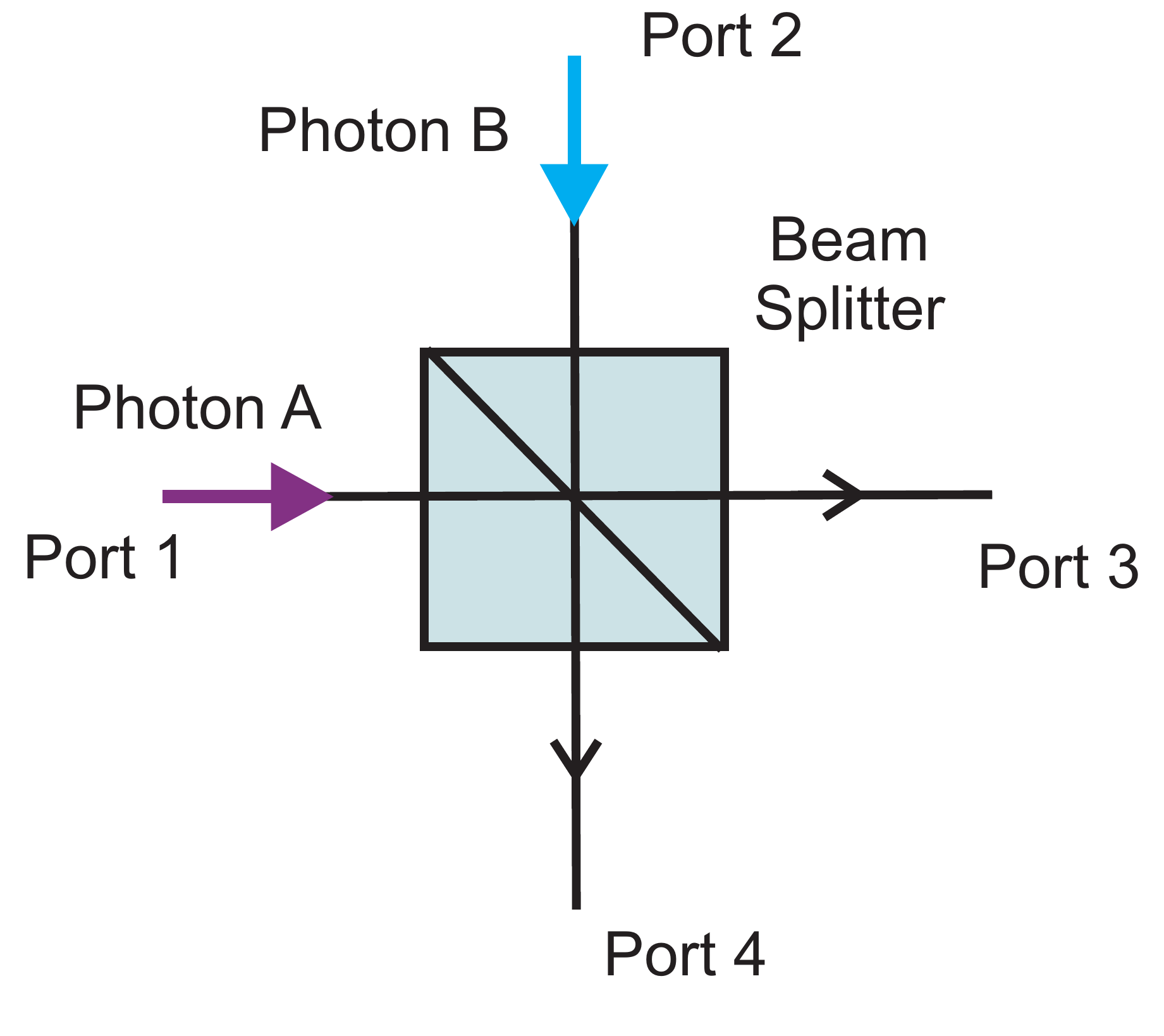}\label{HOMfiga}}
\subfigure[]{
\includegraphics[height=1.2in]{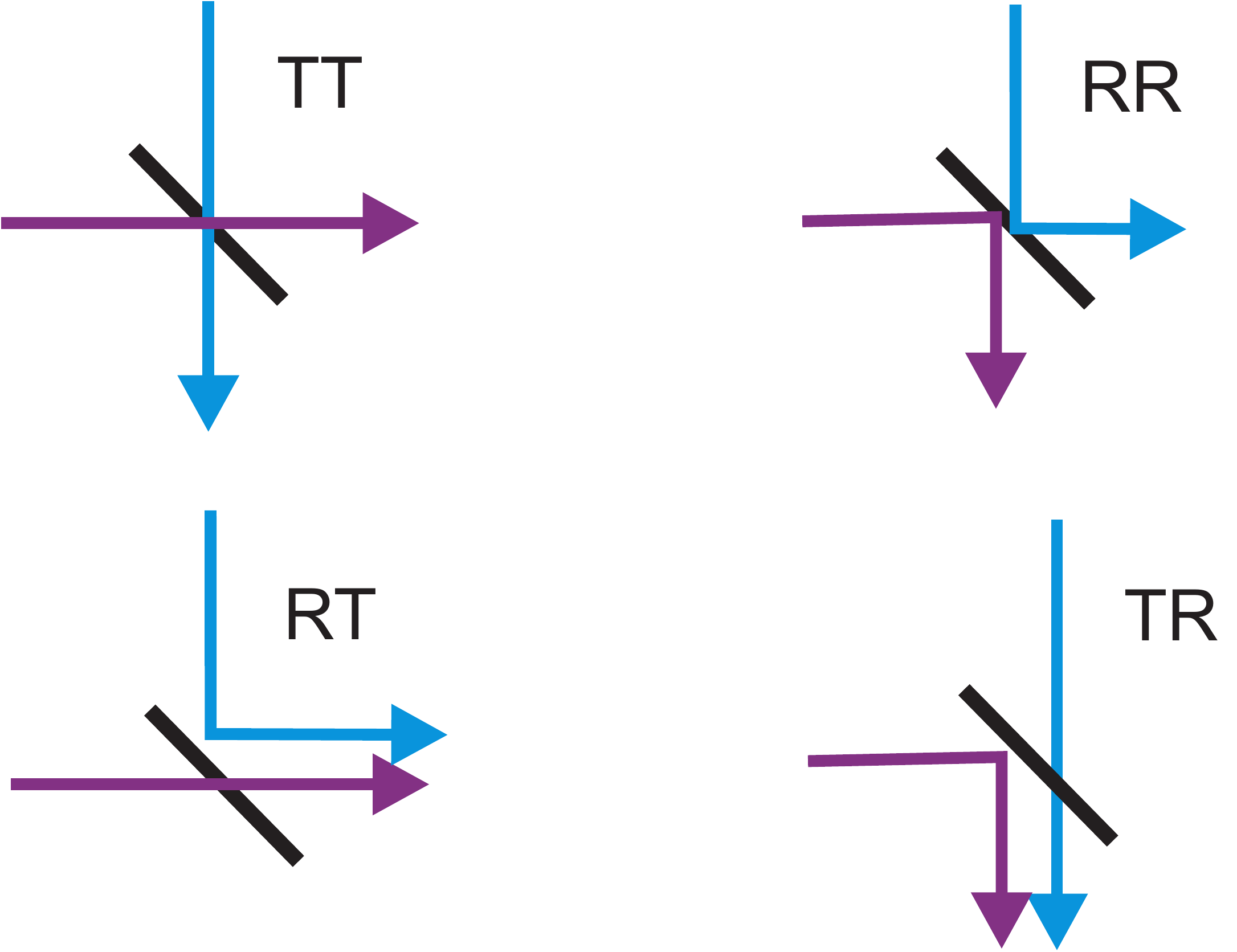}\label{HOMfigb}}
\caption{Two photons entering a beam splitter at different ports (a) lead to four different outcomes (b), labeled according to whether each
photon reflects (R) or transmits (T). Outcomes on the top line (RR and TT) are indistinguishable. Being of equal magnitude
but opposite sign, their amplitudes cancel, so coincidences between output ports vanish. Although exiting at a random output port, both photons are always
found to cluster together in the \emph{same} port.} \label{HOMfig}
\end{center}
\end{figure}

The Hong-Ou-Mandel (HOM) effect \cite{HOM} is probably the best known two-photon interference effect. Two identical photons are
simultaneously incident on different inputs of a 50/50 beam splitter (BS) as in Fig. \ref{HOMfiga}. Each photon could exit either output port, so naively one
expects nonzero amplitudes for three possible outcomes: both exiting at port 3, both exiting at port 4, or one photon each at ports 3 and 4. But in fact,
no coincidences are seen  between 3 and 4; the two photons always leave at the same port. \emph{Which} port the pair exits is entirely random.
Coincidences between the two output ports are absent because of cancelations between the two indistinguishable processes on the upper line of Fig. \ref{HOMfigb}. As a result, the two photons always remain clustered together in the same output spatial mode. This gives a method for measuring
time intervals to sub-picosecond level accuracy: as a delay between the photons varies, the coincidence rate exhibits a sharp dip (the HOM dip) when the wavefunctions briefly overlap on the BS.

\begin{figure}
\begin{center}
\subfigure[]{
\includegraphics[height=1.0in]{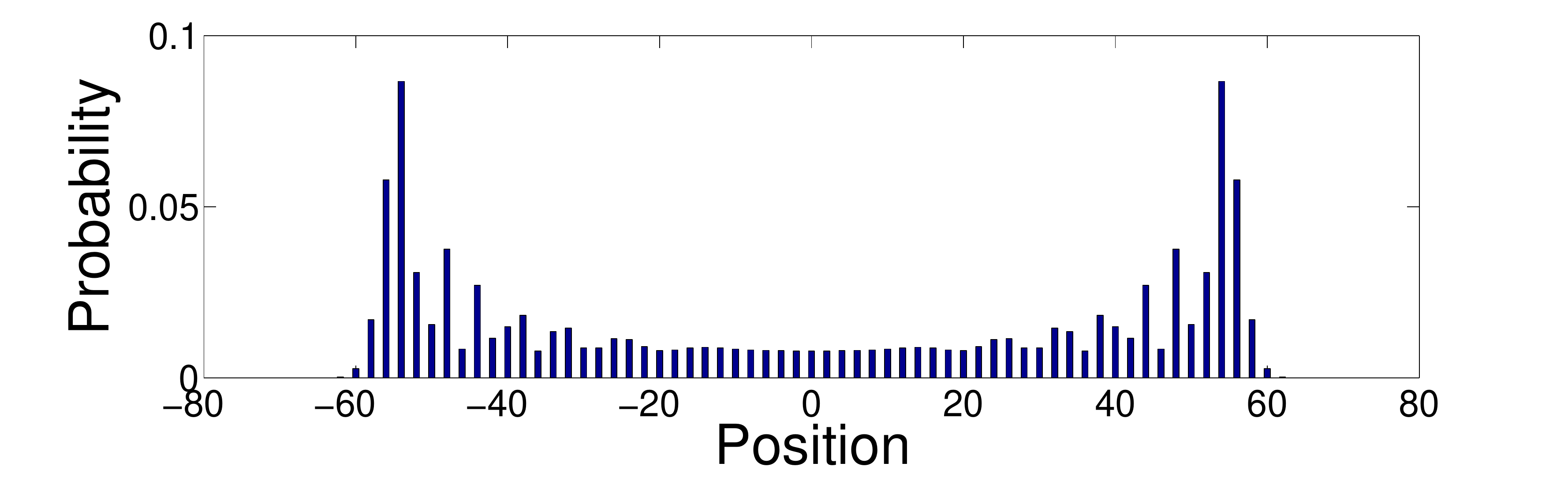}\label{walkfiga}}
\subfigure[]{
\includegraphics[height=1.0in]{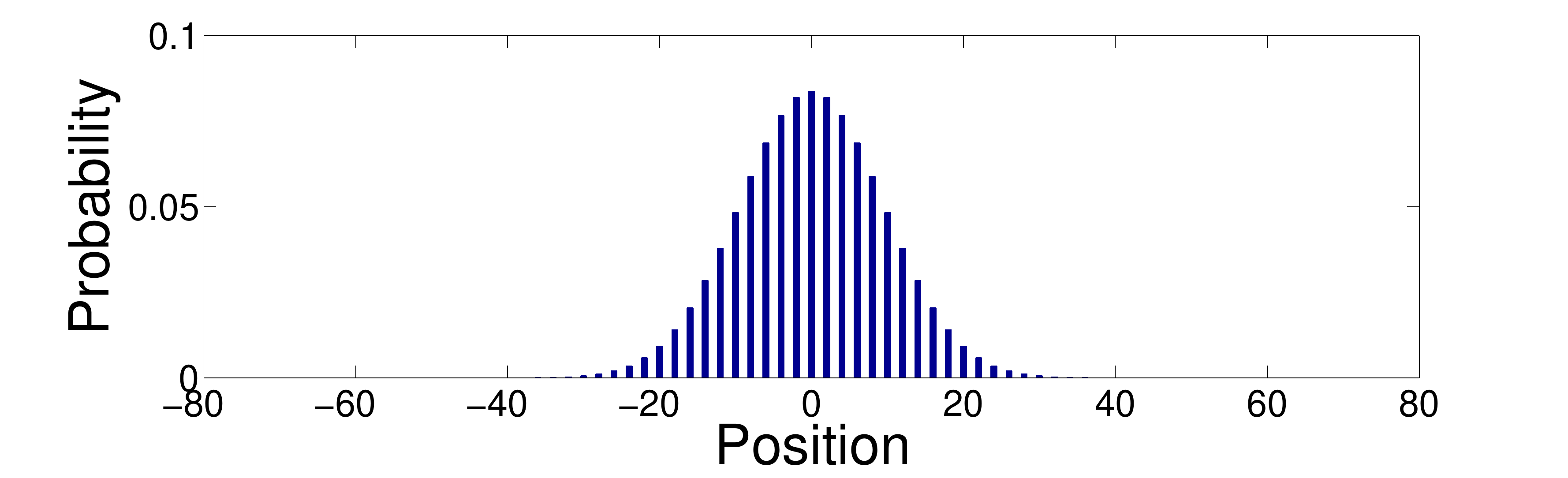}\label{walkfigb}}
\caption{(a) Typical 1-dimensional (Hadamard) quantum walk. Spatial probability distributions have low
probability of remaining near the origin and widths that grow linearly with time. (b) In contrast, classical walks
yield approximately Gaussian distributions, whose widths grow more slowly, $\sim\sqrt{t}$.} \label{walkfig}
\end{center}
\end{figure}

Quantum walks \cite{kempe,ambainis} are currently a subject of
extensive investigation, in part because they have been shown to be formally equivalent to a universal quantum computer \cite{childs,lovett,portugal}, and so they provide new insights into a variety of quantum algorithms, especially quantum search algorithms. In a Hadamard walk, the walker's amplitude produces two peaks that move ballistically in opposite directions (Fig. \ref{walkfiga}), so the distribution's width grows linearly in time, $\sigma\sim t$. In contrast, classical random walks give approximately Gaussian distributions whose widths
spread diffusively ($\sigma\sim \sqrt{t}$).
Consequently, quantum walks can probe large regions faster than classical walks, leading to quantum speedups of
walk-based algorithms. Details differ for specific implementations of quantum walks, but discussion of quadratic speedups can be found in \cite{amb} for Hadamard walks and in \cite{grover,shenvi} for Grover walk-based searches.

%\begin{figure}[t!]
%\centering
%\includegraphics[totalheight=1.4in]{quantum_walk_example}
%\caption{Typical example of a one dimensional quantum walk. The spatial probability distribution peaks at the two ends, with small
%amplitude of remaining near the origin. The distance between the peaks grows linearly with time. In contrast, classical walks yield approximately Gaussian distributions,
%with most amplitude staying near the origin. The width of the classical Gaussian grows more slowly, $\sim\sqrt{t}$.}%\label{walkfig}
%\end{figure}

\begin{figure}
\begin{center}
\subfigure[]{
\includegraphics[height=1.1in]{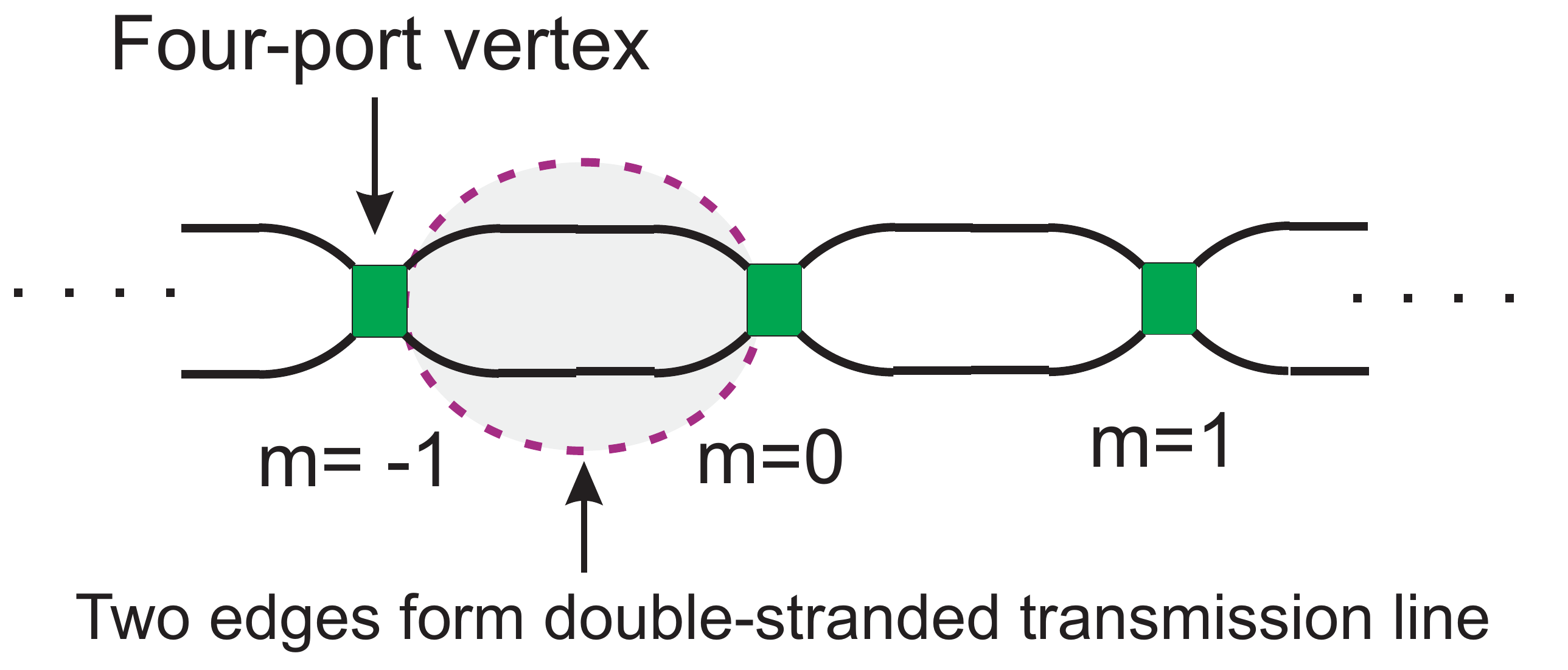}\label{setupfig}}
\subfigure[]{
\includegraphics[height=.8in]{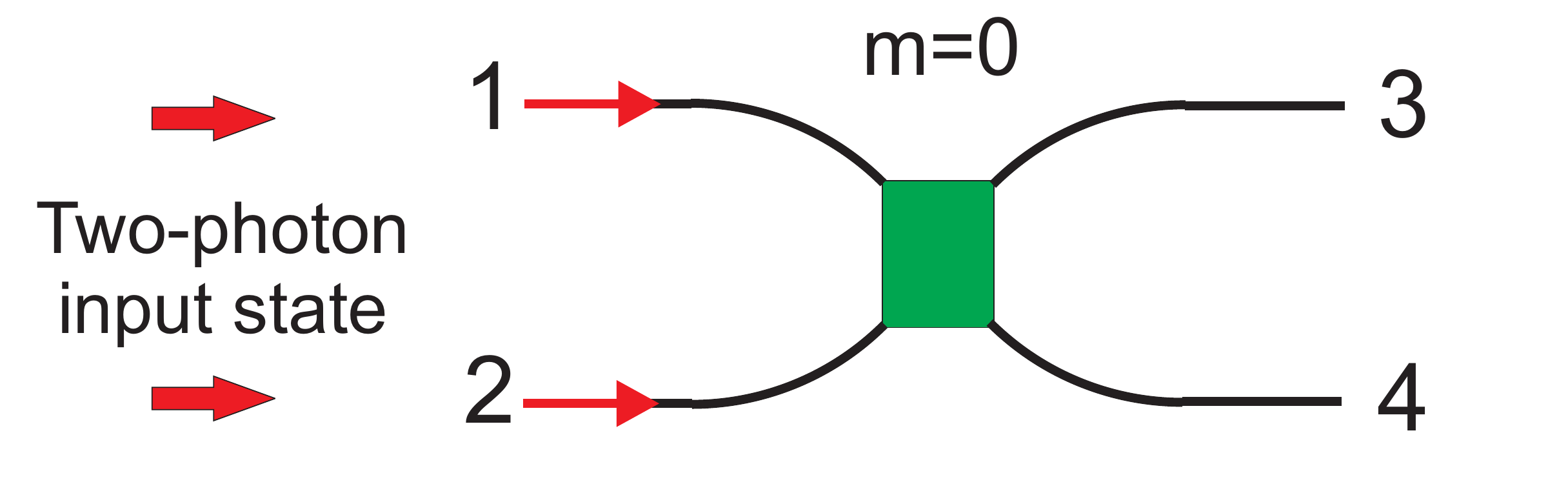}\label{4portafig}}
\caption{(a) A chain of four-ports connected by pairs of edges. Each edge pair is thought of as a single double-stranded connection line,
and vertex positions are labeled by integers. (b) The initial state consists of two indistinguishable, right-moving photons injected into ports 1 and 2 of the $m=0$ multiport, in the middle of the chain.} \label{walk2fig}
\end{center}
\end{figure}

A variety of two-particle quantum walks have been studied \cite{omar,pathak,shaiu,gamble,zahr,peruzzo,stefanak,rohde,berry,carson}. Here we look at a novel
arrangement in which two indistinguishable bosons undergo a discrete-time quantum walk along a dual rail or ladder type system (Fig. \ref{setupfig}), with a
four-dimensional Grover coin at each vertex. Such a chain could be considered as a pair of quantum wires (representing a pair of states), with the Grover coins serving as directional couplers \cite{fan,nikol} between them; more pertinently to our purposes here, we may also think of the system as a single double-stranded quantum wire in which we care only about the horizontal location of the particle, not whether it is on the upper or lower strand.

\begin{figure}[t!]
\centering
\includegraphics[totalheight=1.9in]{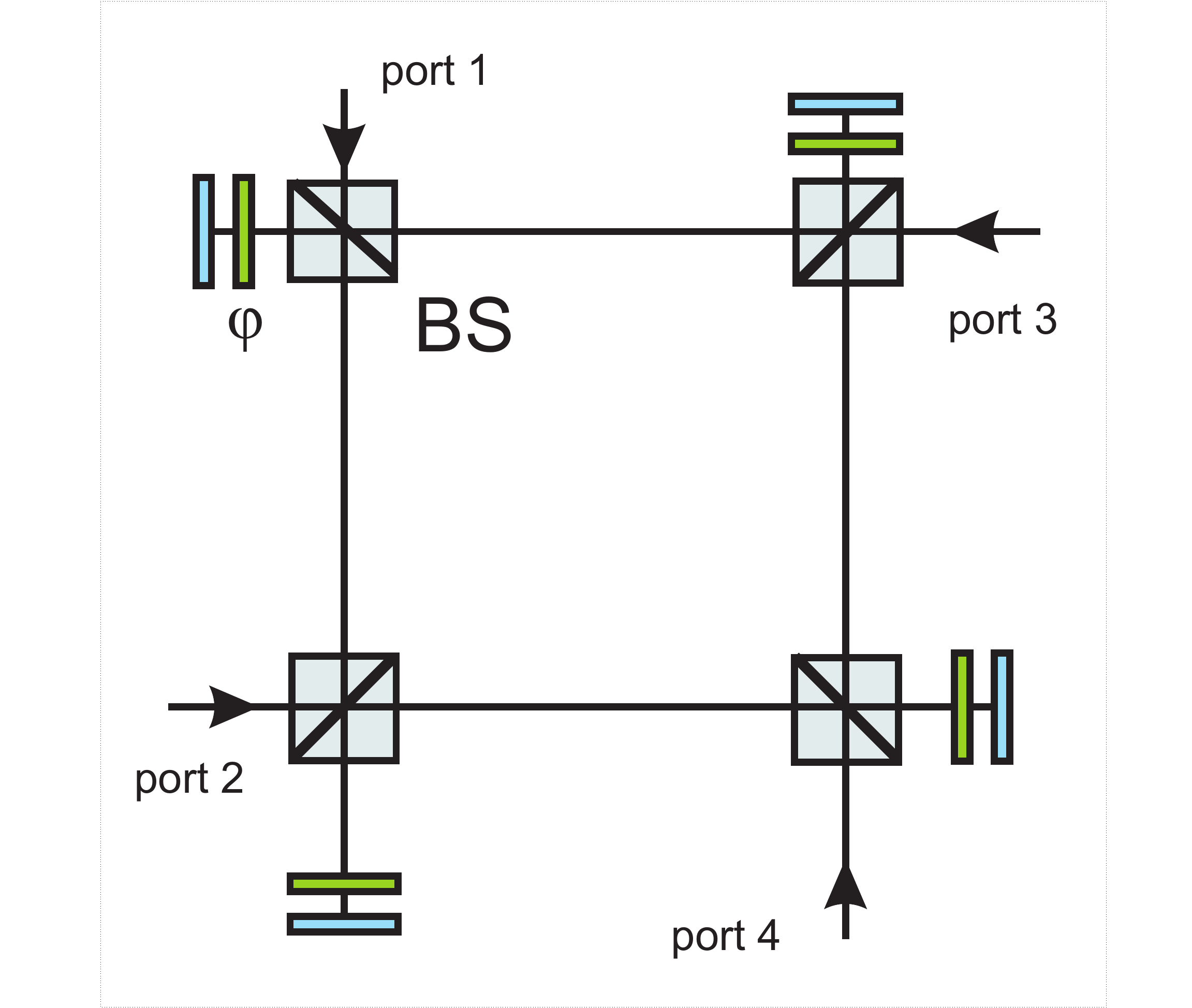}
\caption{One possible optical implementation of the four-dimensional Grover coin, the directionally-reversible four-port device. Photons can both enter and exit each of the four external ports. At each corner of the loop there is a beam splitter, a mirror (green) and a phase shifter (blue). Similar $n$-port devices can be constructed for any $n$; their workings are analyzed in details in \cite{simon1} and the three-port version is experimentally demonstrated in \cite{osawa1}. }\label{roundfig4}
\end{figure}

For specificity, assume that the walking particles are photons. Then the Grover coins can be
implemented using a linear-optical four-port (Fig. \ref{roundfig4}), which is a special case of the directionally-unbiased multiports studied in \cite{simon1,simon2,simon3,simon4,osawa1,osawa2}. The three-port version of this device has been demonstrated in a tabletop set-up  \cite{osawa1}, and initial work on integrated chip versions of such structures is underway. The
system is based entirely on linear optics, with no interactions between the photons. In particular, if the photons are \emph{distinguishable}, then each exhibits an
independent quantum walk and can later be detected in widely separated spatial regions. However, once the photons become \emph{indistinguishable}, the two-photon interference alters the
picture: it is shown below that for a particular input state the two photons remain spatially clustered and are always found at the \emph{same} horizontal location at
each moment. Moreover, the two-photon amplitude shows no sign of the randomness normally associated with random walks: it splits into a quantum superposition of
two localized packets that each move deterministically over time with the two photons always remaining clustered in one packet or the other.

These effects depend only on indistinguishable photons being inserted into the same Grover coin vertex simultaneously; entanglement is not required. However, if the photons \emph{are} entangled then they remain together with entanglement undiminished as they move, opening up new possibilities for quantum information processing, as briefly commented upon in the conclusion.

Experimentally, the most practical realization of the structures described here are on integrated optical chips. Losses, decoherence, and chip imperfections will of course limit the possible walk lengths of experimental implementations. Up to this point, quantum walks of both single photons and of entangled photon pairs have typically been implemented using integrated optics for walks of lengths on the order of five to ten time steps (for example, \cite{peruzzo,sansoni,crespi,su}), although proposals have been made for arrangements that may allow longer walks \cite{geraldi}. In what follows, we assume an idealized system, neglecting losses and other imperfections.

\section{Setup and Main Result.} Consider a directionally-unbiased four-port acting as a Grover coin \cite{moore,carneiro}, with ports labeled as in Fig. \ref{4portafig}. The
action of the four-port is given by the unitary matrix \begin{equation}U={1\over 2}\left( \begin{array}{cccc} -1 & 1 & 1 & 1\\ 1 & -1 & 1& 1\\ 1 & 1 & -1 & 1\\ 1 & 1& 1& -1
\end{array}\right) ,\label{Umatrix} \end{equation} where rows and columns represent the four ports. Regardless of which port a photon enters, exit amplitudes at all outputs are real and equal in magnitude.
Importantly, the amplitude to reflect back to the input port has an extra minus sign relative to all other transitions.

Consider two photons entering the linear chain of Fig. \ref{setupfig} simultaneously. Assume one enters port $1$ and the other enters port $2$ of the same
multiport, as in Fig. \ref{4portafig}, somewhere in the middle of the chain, far enough from the ends that we don't need to worry about the photons leaving the system during the time duration of the experiment.  Experimentally, two photons can be produced simultaneously using spontaneous parametric down conversion \cite{boyd} and then coupled into the chain by means of an electro-optical or magneto-optical switch.
Horizontal positions are specified by an integer $m$ corresponding to the multiport label, and discrete time $t=nT$ by integer $n$, where $T$ is the photon travel time between multiports.  At time $n=0$, the photons are moving rightward, entering the $m=0$
multiport.

Then if the locations of the two photons at any later time $n>0$ are measured, two striking things are found.  First, the photons cluster spatially: if the two parallel
input/output edges between adjacent multiports are treated as a single double-stranded connecting line, then the photons are always found on the \emph{same}
double line. Assuming no loss and ideal measuring devices, measurement at any horizontal location always finds either two photons or none.  This can
be seen as a quantum walk analog of the HOM effect: amplitudes for indistinguishable outcomes in which the photons move apart always cancel among themselves, as will be shown in the next section.
However, the HOM quantum interference effect occurs just once, whereas clustering of the walk persists indefinitely, over an arbitrarily long sequence of steps.

\begin{figure}
\centering
\includegraphics[totalheight=1.9in]{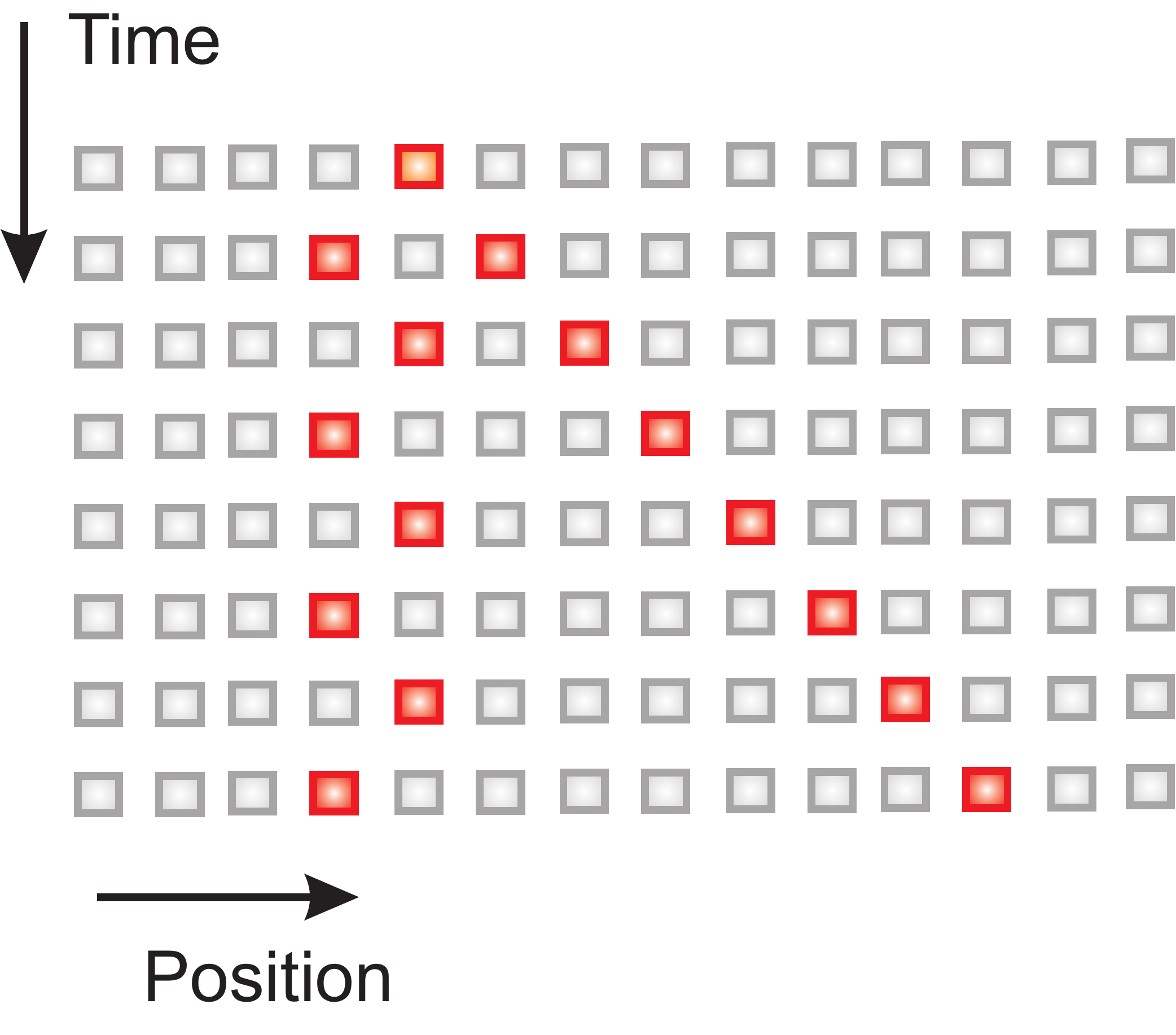}
\caption{After the first step, the amplitude splits: it becomes an equal superposition of two photons moving away ballistically, and two photons
oscillating near the initial point.}\label{splitfig}
\end{figure}

Second, this clustered two-photon amplitude behaves in an unusual manner. It breaks after the first step into a \emph{sustainable superposition of two distinct localized
states} (Fig. \ref{splitfig}). One two-photon cluster in the superposition moves away from the starting point ballistically, exhibiting no randomness. The
other cluster stays near $m=0$, bouncing back and forth between two adjacent locations (a phenomenon dubbed oscillatory localization \cite{amb3} ). So another way to look at the state is as an odd sort of
two-photon clustered Schrodinger cat state, in which the parts of the cat rapidly separate from each other: the two-photon cat speeds away after a rat \emph{and} simultaneously
remains rocking contentedly in its warm cat bed.

For comparison, imagine a \emph{single} photon entering the present system, initially localized on one input port. This can be seen as the sum of two states: one symmetric over the upper and lower lines and one antisymmetric,
\begin{eqnarray}|1\rangle &=& {1\over 2}\left( |1\rangle +|2\rangle \right) +{1\over 2}\left( |1\rangle -|2\rangle \right) \\
&=& |\psi_s\rangle +|\psi_a\rangle .\end{eqnarray}
The symmetric part will always continue rightward at each step, while the antisymmetric portion reflects at each step, leading to oscillations.
%The new feature in the present two-particle case is that both particles \emph{simultaneously} exhibits  \emph{both} of the latter behaviors at once, with both photons always seen to be doing the \emph{same} thing when measured.
These single-particle behaviors have been previously discussed in \cite{amb3}  for  Grover coin systems.  What is remarkable in the two-particle  case is that measurement of the two particles will always find them in the \emph{same} part of the superposition; one will never be found in the oscillating portion and the other in the ballistic portion. Which part of the superposition the two photons are found to be clustered in is completely random, just as the output port in which the two photons are clustered in the HOM effect is purely random.

The two halves of the clustered amplitude do not spread as they propagate, exhibiting soliton-like behavior. Note that the wavepacket spread in quantum walks is of
statistical origin, not a result of dispersive material properties. So cancelation of spreading occurs in the current system from linear
interference processes, with no need for nonlinear interactions.

\section{Time evolution}\label{evolvesection}

\subsection{First time step}

Here we sketch the time evolution of the system. Additional details of the calculations can be found in Appendix A.

Momentarily treating the photons as distinguishable, there are 16 possible exit outcomes from the four-port for the input state of Fig.
\ref{4portafig}. Applying tensor product $U\otimes U$ to the two-photon input, each of these real exit amplitudes has absolute value $\left( {1\over 2}\right)^2 ={1\over 4}$
if the photons exit at \emph{different} ports, or ${\sqrt{2}\over 4}={1\over {2\sqrt{2}}}$ if they exit at the \emph{same} port. (The extra $\sqrt{2}$ appears when the
indistinguishability is restored, due to the normalization of two-boson Fock states, $|2\rangle ={1\over \sqrt{2}}\left( a^\dagger\right)^2|0\rangle$.)
Amplitudes gain one minus sign for each photon that exits back out the port through which it entered. So signs and magnitudes of all transition amplitudes are readily
obtained; these are tabulated in the Appendix.
%Turning to the signs of those amplitudes, these are readily seen to be given by the $+$ and $-$ signs in the table of Fig. \ref{signfig}.

%
%\begin{figure}[h!]
%\centering
%\includegraphics[totalheight=1.5in]{signs}
%\caption{The signs of the real exit amplitudes for all processes starting from the initial state of Fig. \ref{4portafig} after one trip through the fourport. }\label{signfig}
%\end{figure}

The initial state is $|\psi_0\rangle_{in} = |12;0,RR\rangle$; the notation $|ij;m,RR\rangle$
(or $|ij;m,LL\rangle$) means one right-moving (left-moving) photon in port $i$ and one in port $j$ at lattice site $m$.
The resulting output state is  \begin{eqnarray}|\psi_1\rangle &=&{1\over {2\sqrt{2}}}\left[ |33;0,RR\rangle + |44;0,RR\rangle -|11;0,LL\rangle  \right. \label{stateeq} \\
& & \; \left. -|22;0,LL\rangle   \right] +{1\over 2}\left[    |34;0,RR\rangle + |12;0,LL\rangle  \right] . \nonumber \end{eqnarray}  Assume we only want to know the exit
direction of the photons (left or right), and don't care if the photon is in the upper or lower channel. Then, clustering can already be seen in the transition
probabilities for the first step of the walk:

$\bullet$ One possible outcome is for both photons entering the left side of the multiport to exit back on the left side ($LL\to LL$).  The probability of this is the sum of three terms corresponding respectively to the amplitudes of both photons exiting at port 1, one photon at each port, and both at port 2:
\begin{eqnarray}P(LL\to LL) &=& P(|12\rangle \to |11\rangle ) + P(|12\rangle \to |12\rangle ) \nonumber \\ & & \qquad\quad  +P(|12\rangle \to |22\rangle ) \\
&=& \left(  -{1\over {2\sqrt{2}}}\right)^2 +\left( {1\over 2}\right)^2 + \left( -{1\over {2\sqrt{2}}}\right)^2\nonumber\\ &=& {1\over 2}.
\label{eq1}\end{eqnarray}
%These are the cases in the four boxes of the lower left quadrant of Fig. \ref{signfig}.
%(Although we don't care if the
%exit is at the upper or lower ports, the cases of both exiting at port 1 and both exiting at port 2 are distinguishable, so their probabilities (not their
%amplitudes) must be added.)

$\bullet$ Similarly, both photons can exit right (ports $3$ and $4$): \begin{eqnarray}P(LL\to RR)
&=& P(|12\rangle \to |33\rangle ) + P(|12\rangle \to |34\rangle ) \nonumber \\ & & \qquad\quad  +P(|12\rangle \to |44\rangle ) \\
&=&\left( +{1\over {2\sqrt{2}}}\right)^2+ \left( {1\over 2}\right)^2
+\left({1\over {2\sqrt{2}}}\right)^2 \nonumber\\ &=& {1\over 2}.\label{eq2}\end{eqnarray}

$\bullet$ Finally, one photon can exit left and one right. The appearance of extra minus signs in half the amplitudes (see Appendix A) leads to complete cancelation: \begin{eqnarray}P(LL\to
LR)&=& P(12\to 13)+P(12\to 14)  \\ & & \qquad  +P(12\to 23)+P(12\to 24)\nonumber\\ &=& 0.\label{eq3} \end{eqnarray}

%
%So now that the magnitudes of all of the possible two-photon exit amplitudes are known, we must look at the signs of those amplitudes.
%Recall that $U$ does not change the phase of the output, except for a reversal of sing for those photons that exit back out the original input port. Therefore the signs are
%given by the purple $+$ and $-$ signs tabulated in the table of Fig. \ref{signfig}.
%
%
%\begin{figure}
%\centering
%\includegraphics[totalheight=2.0in]{signs}
%\caption{Signs of all possible two-photon exit amplitudes from the four-port, ro given input state $|\psi_o\rangle$.}\label{signfig}
%\end{figure}

The result is that \emph{even though the photons do not interact and should walk independently, they in fact always step in the same direction: both go right or
both go left}. Destructive interference between indistinguishable amplitudes conspires to eliminate outcomes in which they step in opposite directions.

\subsection{Subsequent steps} The paragraphs above describe the first step. Transition amplitudes can again be tabulated to find the outcomes of subsequent steps.
Summing over unmeasured intermediate states in previous steps, one finds the amplitude splitting into an equal superposition of two two-photon states.
%For example, after the first step the $(1,1)$ and $(2,2)$ inputs will
%always be likely in subsequent steps (due to symmetry), and the amplitudes for the transitions $(1,1)\to (1,3)$ and $(2,2)\to (1,3)$ are indistinguishable; so
%their amplitudes will cancel when added.)

%\begin{figure}[h!]
%\centering
%\includegraphics[totalheight=1.3in]{4portb}
%\caption{}\label{4portbfig}
%\end{figure}
The output of the first step (Eq. \ref{stateeq}) can be written as \begin{equation} |\psi_1\rangle_{out} ={1\over \sqrt{2}} \left( |\psi_t;0,RR\rangle +|\psi_r;0,LL\rangle\right)  ,\end{equation} where \begin{eqnarray} & & |\psi_t;m,R\rangle =
{{|33;m,RR\rangle +|44;m,RR\rangle
}\over 2}  \\ & & \qquad \qquad \qquad + {{|34;m,RR\rangle}\over \sqrt{2}}\nonumber \\
 && \;=  {{|11;m+1,RR\rangle +|22;m+1,RR\rangle
}\over 2} \label{psit} \\ & & \qquad \qquad \qquad + {{|12;m+1,RR\rangle}\over \sqrt{2}} \nonumber \\
& &  |\psi_r;m,LL\rangle = -{{ |11;m,LL\rangle +|22;m,LL\rangle }\over 2}\label{psir} \\ & & \qquad \qquad\qquad + {{|12;m,LL\rangle}\over \sqrt{2}}\nonumber \\
 && \;=   -{{ |33;m-1,LL\rangle +|44;m-1,LL\rangle }\over 2}\\ & & \qquad \qquad\qquad + {{|34;m-1,LL\rangle}\over \sqrt{2}}\nonumber
.\end{eqnarray} Here, we used the fact that states leaving ports 3 and 4 enter the adjacent vertex at ports 1 and 2, respectively.
%
%The states $|\psi_t;m,R\rangle$ and $|\psi_r;m,L\rangle$, along with their time-reversed versions, are the basic building blocks for the full dynamics of the
%system.

Suppressing some labels for brevity, one finds that applying $U\times U$ again gives \begin{eqnarray} &  {{|11\rangle +|22\rangle}\over 2} \to  {{  |33\rangle
+|44\rangle +|11\rangle
+|22\rangle}\over 4}  + {{  |34\rangle -|12\rangle }\over  {2\sqrt{2}}} & \label{evolve1}\\
& {{|12\rangle}\over \sqrt{2}}  \to   {{ |33\rangle +|44\rangle -|11\rangle -|22\rangle }\over 4} + {{ |34\rangle +|12\rangle }\over {2\sqrt{2}}} .&
\label{evolve2}\end{eqnarray} Taking the sum of these as in Eq. \ref{psit}, one finds that the amplitudes $|11\rangle,$ $|22\rangle$, and $|12\rangle$  cancel out at each step, so that $|\psi_t\rangle $ simply reproduces itself, but shifted one step to the right:
\begin{equation} |\psi_t;m,RR\rangle \to |\psi_t;m+1,RR\rangle \to |\psi_t;m+2,RR\rangle \to \dots .\end{equation} This is the \emph{ballistic} state: it is \emph{totally transmitting} at
each step. If the initiating state of the walk had been moving left ($|34;m,LL\rangle$ instead of $|12;m,RR\rangle$), similar ballistic motion occurs to the left.

%\to {1\over 2}\left( |33\rangle +|44\rangle \right) +{1\over \sqrt{2}} |34\rangle $ and
%$|\psi_r\rangle  \to -{1\over 2}\left( |11\rangle +|22\rangle \right) +{1\over \sqrt{2}} |12\rangle $.  It is then seen that over one time step $|\psi_t;m,R\rangle$ is simply %reproduced, but shifted one step to
%the right: $|\psi_t;m,R\rangle \to |\psi_t;m+1,R\rangle \to |\psi_t;m+2,R\rangle \to \dots .$

The multiport action on $|\psi_r\rangle$ of Eq. \ref{psir} is found by taking the difference of Eqs. \ref{evolve1} and  \ref{evolve2}, leading to cancellation of $|33\rangle$, $|44\rangle$, and $|34\rangle$ terms. So $|\psi_r;m,RR\rangle$ simply reflects at each multiport encounter, causing it to bounce back and forth indefinitely: \begin{equation} |\psi_r;m,RR\rangle \to
|\psi_r;m-1,LL\rangle\to |\psi_r;m,RR\rangle \to \dots\end{equation} The state is \emph{totally reflecting} at each step, and the amplitude acts as if it is confined in a virtual cavity, oscillating between lattice sites $m=0$ and $m=1$.

Up to a spatial shift of one step per unit time, the states $|\psi_t\rangle$ and $|\psi_r\rangle$ are both eigenstates of $\left( U\times U \right)^2$, with eigenvalue $+1$, so evolution on subsequent steps is simply a repetition of what happened in the first two steps: one two-photon amplitude repeatedly reflects, the other repeatedly transmits.

So the photons remain spatially clustered as they walk along the line.   This quantum walk-based analog of the HOM effect might be referred to as a
\emph{quantum-clustered two-photon walk}. But in addition, the two-photon state at each moment localizes onto a quantum superposition of just two nondispersive spatial
amplitudes: one moves ballistically at constant speed, while the other flips direction at each step and never moves more than one unit from its starting point. This is analogous to the single-particle Grover walk behavior, but with the unexpected feature that the two photons are always found clustered in the same localized part of the distribution and never separate from each other.
%\emph{Although
%each photon individually exhibits a standard quantum walk, the two-photon amplitudes exhibit behavior that seems to %retain no random walk quality. }

%\pagebreak
%
%\begin{figure}[h!]
%\centering
%\includegraphics[totalheight=1.8in]{measure}
%\caption{Measurements are made on pairs of edges, such as the two intersecting the dotted red line. }\label{measfig}
%\end{figure}

%
%In pictorial form, this means:
%
%
%\begin{figure}[h!]
%\centering
%\includegraphics[totalheight=1.6in]{output}
%\caption{}\label{outfig}
%\end{figure}

\begin{figure}
\centering
\includegraphics[totalheight=2.7in]{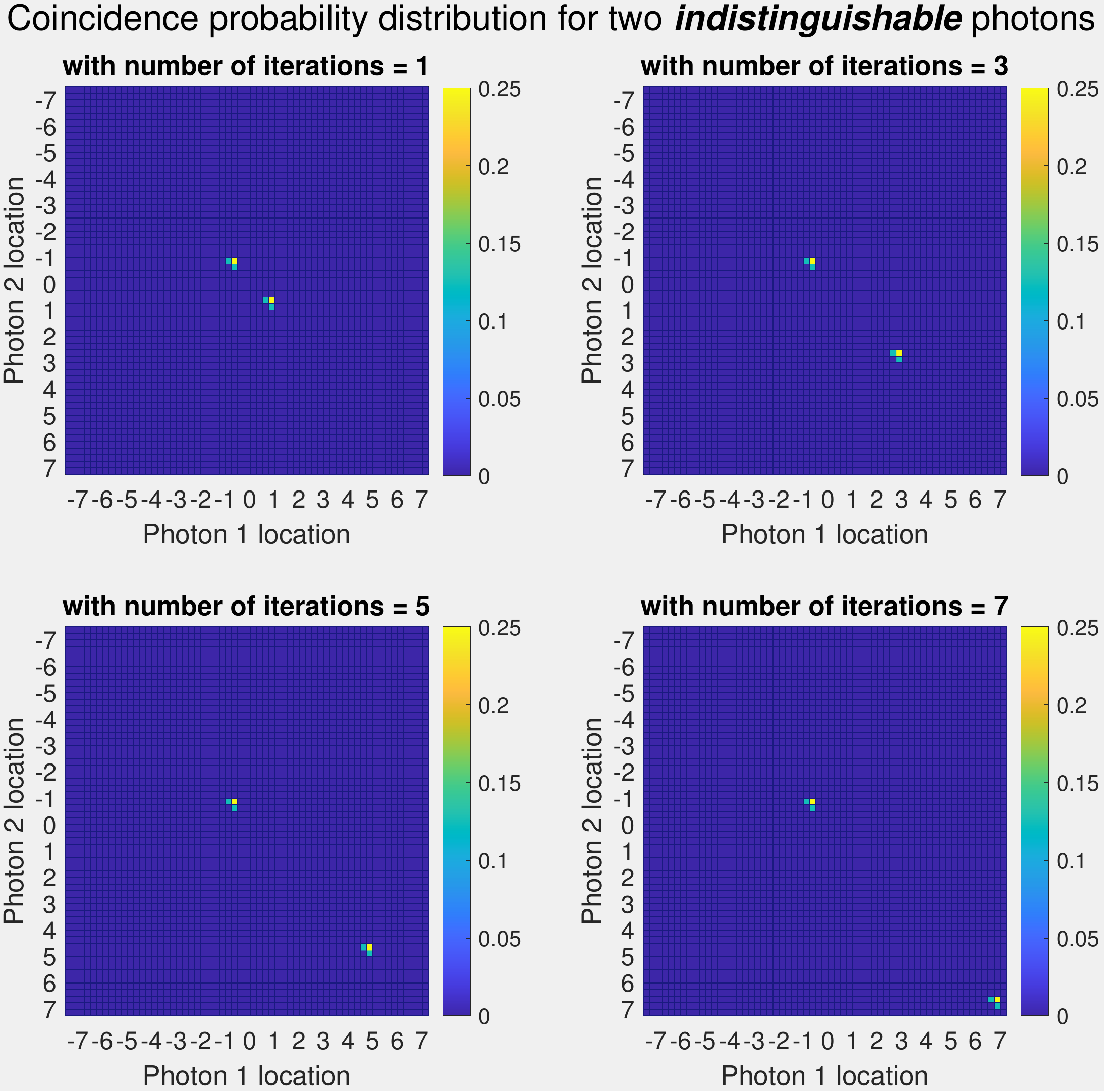}
\caption{The spatial distribution of the photon amplitudes for two photons at four different times, given an initial state with two \emph{indistinguishable} photons entering ports one and two of the four-port at position $m=0$. The two axes give the locations of the two photons, labelled by the integer-valued four-port index. It is seen that the amplitude splits into two localized components, but that the two photons are always found clustered together within the \emph{same} component, as indicated by the absence of amplitude away from the descending diagonal.  }\label{indistfig}
\end{figure}

\begin{figure}
\centering
\includegraphics[totalheight=2.7in]{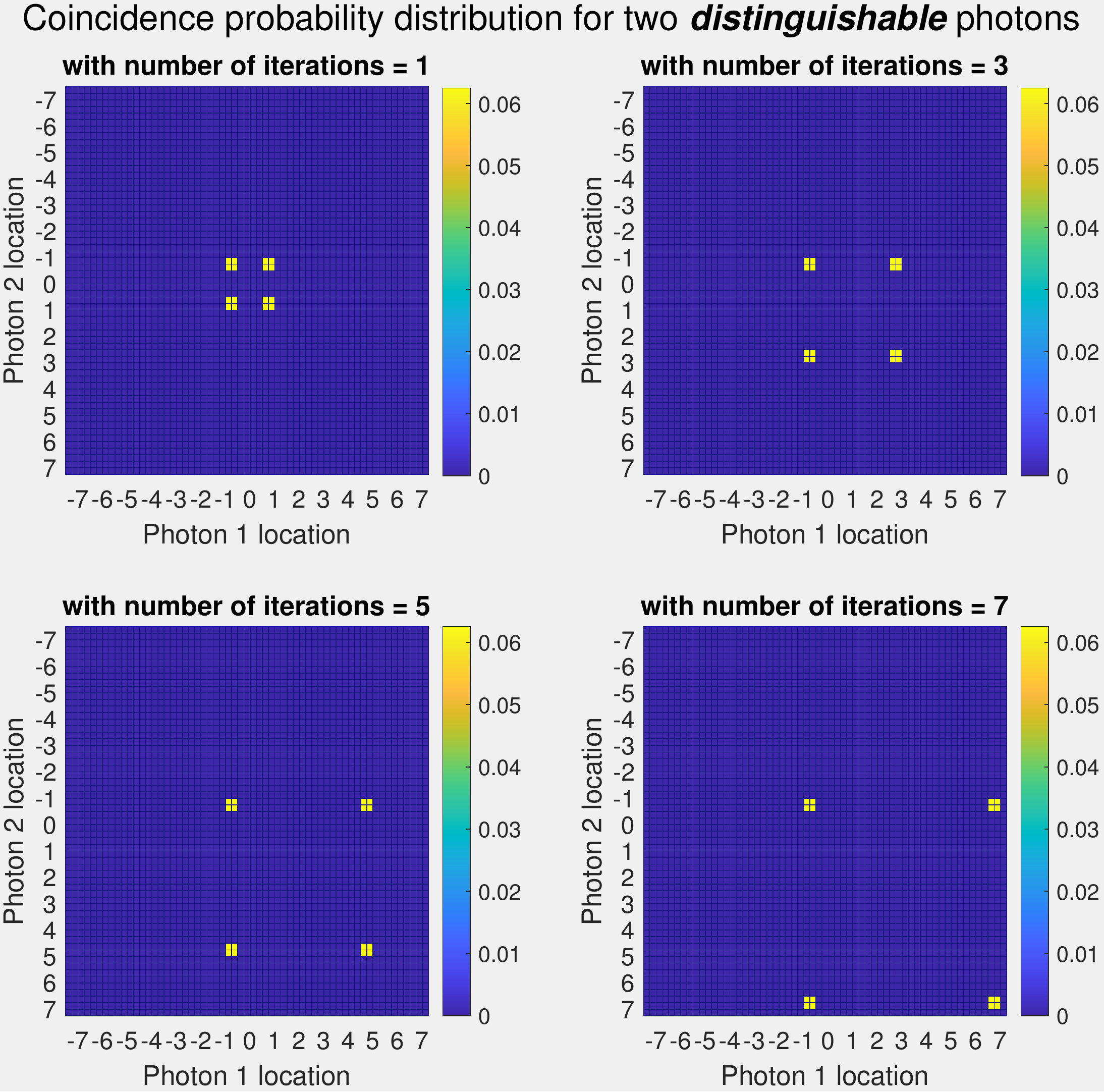}
\caption{The spatial distribution of the photon amplitudes for two \emph{distinguishable} photons. In contrast to the indistinguishable case, off-diagonal terms appear, indicating that the two photons no longer remain clustered together.}\label{distfig}
\end{figure}

The behavior of the system is shown in Fig. \ref{indistfig}, where the amplitude for the position of each photon is shown. It is clearly seen that the amplitude splits into two localized portions, with one portion staying near the origin and the other moving away at constant speed. Moreover, it can be seen that the two indistinguishable photons remain together: there is no amplitude away from the diagonal. In contrast, if the two photons are distinguishable (Fig. \ref{distfig}), the lack of destructive interference leads to the appearance of nonzero off-diagonal amplitudes, indicating that the photons may become spatially separated.

In all of the considerations above, the two photons were assumed to be in a product states. In the next section, we examine the behavior of polarization entangled states in this system.

\section{Polarization-Entangled Input}\label{entsection}
Up to now, no entanglement has been assumed between the two input photon states.  Now suppose that the two photons inserted at the origin are polarization-entangled. In particular, define the states:
\begin{eqnarray}
|A_\pm^{ij}\rangle &=& {1\over \sqrt{2}} \left\{|iH\rangle |jV\rangle  \pm |iV\rangle |jH\rangle\right\} \\
|B_\pm^{ij}\rangle &=& {1\over \sqrt{2}} \left\{|iH\rangle |iV\rangle  \pm |jV\rangle |jH\rangle\right\}
\end{eqnarray}
Here, $i,j$ label the input/outport ports, while $H$ and $V$ label polarization states at each port. When
necessary, we can add position and direction labels, for example $|A_\pm^{ij},m,LL\rangle$ or $|B_\pm^{ij},m,LR\rangle$. These
states are polarization-entangled for $i\ne j$. For $i=j$, the $+$ states are product states, while the $-$ states vanish. It is straightforward to work out the action of the multiport on these states by the same means as in section \ref{evolvesection}.

Analogous to previous sections, focus on the initial state $|\psi_0\rangle_{in} = |A_+^{12},m\rangle.$ (Additional analysis of the $A$ and $B$ states, along with a third set of polarization-entangled states is given in Appendix B.) The output of the first step is
\begin{eqnarray}|\psi_0\rangle_{out}& =& {1\over 2}\left( \right. |A_+^{12},m,L\rangle+|A_+^{34},m,R\rangle \\ & & \qquad \quad
+|B_+^{34},m,R\rangle-|B_+^{12},m,L\rangle\left. \right)_{out} .\nonumber \end{eqnarray} It is readily seen that once again the photons cluster: both exit
left (ports $1$ and $2$) or both exit right ($3$ and $4$). \emph{Crucially for entanglement, the two photons are always at different ports: one exits onto the upper line and one onto the lower line.}

This state enters adjacent multiports at the next step as
\begin{eqnarray}|\psi_1\rangle_{in}& =& {1\over 2}\left(
|A_+^{34},m-1,LL\rangle+|A_+^{12},m+1,RR\rangle \right. \\ & & \qquad  + \left. |B_+^{12},m+1,RR\rangle-|B_+^{34},m-1,LL\rangle\right)_{in}, \nonumber \\
& =& {1\over \sqrt{2}}\left( |\psi_r,m-1,LL\rangle+|\psi_t,m+1,RR\rangle\right)  ,\end{eqnarray} where
%\begin{eqnarray}|\psi_r\rangle &\equiv& {1\over
%\sqrt{2}}\left( |A_+^{34}\rangle -|B_+^{34}\rangle \right) \\  |\psi_t\rangle &\equiv& {1\over \sqrt{2}}\left( |A_+^{12}\rangle +|B_+^{12}\rangle
%\right)\end{eqnarray}
\begin{equation}|\psi_r\rangle \equiv {1\over
\sqrt{2}}\left( |A_+^{34}\rangle -|B_+^{34}\rangle \right)\end{equation} and \begin{equation}|\psi_t\rangle \equiv {1\over \sqrt{2}}\left( |A_+^{12}\rangle +|B_+^{12}\rangle
\right) .\end{equation}
[Similarly, if the input state had been $|B_+^{12},m\rangle$, the resulting output would have been  ${1\over \sqrt{2}}\left( |\psi_r,m-1,LL\rangle -|\psi_t,m+1,RR\rangle\right)
$.] Calculations similar to the previous section show these again to be totally transmitting and totally reflecting:
$|\psi_r\rangle $ subsequently oscillates back and forth near the origin, while $|\psi_t\rangle$ remains unchanged aside from repeatedly shifting rightward by one step per unit time. So the picture
remains the same as in the previous section, but with the additional feature that \emph{the polarization-entanglement remains undiminished as the walk proceeds}. Which polarization is on which edge is indeterminate until measured, and clustered two-photon states remain polarization-entangled between
upper and lower edges as they propagate. Perfect state transport (PST) occurs probabilistically ($50\%$ probability) for $|\psi_t\rangle$, delivering two entangled photons
to the same horizontal location at the same time.

\section{Coherent state input}

It is natural to ask how other types of states propagate through this system.  Consider, for example, coherent state input. In particular, consider the state \begin{equation}|\psi_{in}\rangle = |\alpha\rangle_1 |\alpha\rangle_2 |0\rangle_3 |0\rangle_4\end{equation} with equal amplitude coherent states entering ports $1$ and $2$ at some lattice site (here we omit directional and site labels again to streamline notation). This state can be written in terms of displaced vacuum states as
\begin{eqnarray}|\psi_{in}\rangle &=& e^{-{1\over 2}|\alpha|^2} e^{\alpha \hat{a}_1^\dagger -\alpha^\ast \hat{a}_1} e^{-{1\over 2}|\alpha|^2} \\
& & \times e^{\alpha \hat{a}_2^\dagger -\alpha^\ast \hat{a}_2}
|0\rangle_1 |0\rangle_2 |0\rangle_3 |0\rangle_4 .\nonumber
\end{eqnarray} After passage through the four-port, the output state is
\begin{eqnarray}|\psi_{out}\rangle &=& e^{-{1\over 2}|\alpha|^2} e^{{1\over 2}\alpha (-\hat{a}_1^\dagger +\hat{a}_2^\dagger +\hat{a}_3^\dagger\hat{a}_4^\dagger)}e^{
-\alpha^\ast (-\hat{a}_1 +\hat{a}_2 +\hat{a}_3+\hat{a}_4)}\nonumber \\ & & \qquad \times e^{-{1\over 2}|\alpha|^2}e^{{1\over 2}\alpha (\hat{a}_1^\dagger -\hat{a}_2^\dagger +\hat{a}_3^\dagger +\hat{a}_4^\dagger)} \\ & & \qquad \times e^{
-\alpha^\ast (\hat{a}_1 -\hat{a}_2 +\hat{a}_3+\hat{a}_4)}
|0\rangle_1 |0\rangle_2 |0\rangle_3 |0\rangle_4 \nonumber\\ &=& e^{-|\alpha|^2} e^{\alpha (\hat{a}_3^\dagger +\hat{a}_4^\dagger) -\alpha^\ast (\hat{a}_3 +\hat{a}_4)}|0\rangle_1 |0\rangle_2 |0\rangle_3 |0\rangle_4\\ &=& |0\rangle_1 |0\rangle_2 |\alpha \rangle_3 |\alpha\rangle_4 . \end{eqnarray}
Iterating the process, it is clear that this balanced double coherent state propagates indefinitely without reflection.

The unidirectional propagation follows from the fact that the amount of amplitude reflected backward along line 1 (for example) contains equal contributions from the light that had entered at ports 1 and 2, but these two amplitudes will reflect out port 1 with opposite sign. Equivalently, the Grover four-port selects out only the portion of the reflected state that is antisymmetric under interchange of upper and lower lines, but the input contained only a symmetric part.

Coherent state amplitudes are inherently fluctuating objects, so arranging equal amplitudes in the two lines over multiple steps may seem unlikely. It is therefore sensible to look at what happens when the input coherent states are unbalanced, with amplitude $\alpha$ in the upper line and amplitude $\beta$ in the lower line: \begin{equation}|\psi_{in}\rangle = |\alpha\rangle_1 |\beta \rangle_2 |0\rangle_3 |0\rangle_4 .\end{equation}
Manipulations along the same lines as above lead after one step through the Grover four-port to the output
\begin{equation}|\psi_{out}\rangle = |{{\beta -\alpha}\over 2}\rangle_1 |{{\alpha -\beta}\over 2}\rangle_2 |{{\alpha +\beta}\over 2}\rangle_3 |{{\alpha +\beta}\over 2}\rangle_4 .\end{equation} There is now both rightward transmission and leftward reflection; however notice that the amplitudes on the two right-moving lines are of equal amplitude again. Thus, any fluctuations in the input amplitude on the left are automatically evened out in the rightward traveling transmitted amplitudes. This is again due to the fact that the two rightward amplitudes are each equal superpositions of the two input modes, so any fluctuation in one input is equally shared between the two outputs.  If measurements are made at some point $N$ steps to the right of the input, the output arriving at a given time is guaranteed to have equal amplitudes in both lines, regardless of reflections or losses, or of initially unequal inputs. A similar conclusion will hold for output measured at some distance to the \emph{left} of the initial point.

\section{Conclusions} We have shown that two-particle quantum
interference allows two non-interacting, indistinguishable walkers to remain clustered as they walk along a chain of directionally-unbiased Grover four-ports. The resulting state is a superposition of two spatially-localized two-photon clusters, one confined near the origin,  the other moving monotonically away.
If the particles are entangled, the pair moves as a single unit, with undiminished entanglement. Whereas perfect state transport  has been demonstrated for single photon states in
various systems \cite{bellec,perez,chapman},  this system demonstrates the existence of PST for entangled multi-particle states as well, with
$50\%$ arrival probability.

Potential applications are readily envisioned.  Entangled pairs can be delivered in a controllable manner to distant locations for standard applications like entanglement swapping, quantum repeaters, or to
control the flow of entanglement for two-photon interference effects in quantum networks \cite{mcmillan}. Since the ballistic part moves with constant speed,
these locations are addressable simply by waiting the appropriate amount of time for the amplitude to arrive. Similarly, new communication or quantum key
distribution protocols can be imagined requiring two spatially-separated participants to simultaneously share equal access to the amplitudes of both photons in
an entangled pair; the presence of two photons in each spatially-separated state ($|\psi_t\rangle$ and $|\psi_r\rangle$) could be used for error detection, for example.
%Similarly, looping the chain around to bring the two spatially-separated amplitudes back together again, new possibilities for two-photon interferometry effects appear, to provide two-particle probing around the resulting ring.

One reason why utilizing the clustering effect can be useful in such applications, rather than simply sending a photon pair along a fiber or through free space, is that by adding phase shifts in the lines between the multiports the flow of the photon pairs can be controlled; they can be stopped at a desired location (oscillating between two adjacent multiports) or their direction of motion can be reversed. This sort of control is something that can't by done with a simple optical fiber, and here it can be done without damaging any entanglement between the photons. The means of such control is readily seen: inserting phase shifts of $\pi\over 2$ to an upper line and $-{\pi\over 2}$ to the corresponding lower line converts the reflecting and transmitting states of Eqs. \ref{psit} and \ref{psir} into each other, allowing the experimenter to controllably switch back and forth between ballistic and oscillating behavior.

Furthermore, the fact that there are two spatially-separated two-photon amplitudes means that those amplitudes can be brought back together and interfered with each other. This provides a means of probing the region that the ballistic portion has traveled through, allowing new two-photon sensing methods. For example, if one photon is vertically-polarized and one horizontally-polarized (the entangled $|A^{12}_+\rangle$ state, for instance), they may gain different polarization-dependent phase shifts due to external magnetic fields; even very small relative phase shifts will destroy the clustering effect, leading to a dip in the two-photon interference (similar to the Hong-Ou-Mandel dip) when the two halves of the superposition are brought back together, allowing sensitive detection of external fields in the relevant region.

In addition, the results of the previous section show that this arrangement can be used to correct for differential losses between pairs of coherent state beams. This is a significant benefit in coherent state interference experiments.

Applications and further properties of this system will be examined in detail elsewhere.

\acknowledgments{This research was supported by the National Science Foundation EFRI-ACQUIRE grant no. ECCS-1640968, AFOSR grant no. FA9550-18-1-0056, and by the Northrop Grumman NG Next. }

\appendix\section{Calculation of Output Probabilities}

Here, for reference, additional details of the calculations used in the main text are filled in. We work in the basis $\left\{ |1\rangle , |2\rangle , |3\rangle , |4\rangle \right\}$, where $|n\rangle$ represents a photon in port $n$. We suppress directional labels when the direction (ingoing or outgoing from the multiport) is clear, as well as dropping site labels when they are not needed.  A single photon state entering or exiting a given multiport can be written as a column matrix: \begin{equation}\left( \begin{array}{c} a \\ b\\ c\\ d \end{array}\right) =a |1\rangle +b|2\rangle +c |3\rangle +d |4\rangle .\end{equation} The matrix $U$ of Eq. \ref{Umatrix} then acts on such column matrices.

For input state $|1\rangle$, transition matrix $U$ then gives output \begin{equation}|1\rangle \to \left({1\over 2}\right) \left( -|1\rangle +|2\rangle +|3\rangle +|4\rangle \right) ;\end{equation} the action on input states $|2\rangle$, $|3\rangle$, $|4\rangle$ are just cyclical permutations of this.

The action on two-particle input states is obtained by taking products of two single particle output states. For example, given a two-photon input state $|12\rangle =|1\rangle |2\rangle $, with one photon entering port $1$ and one photon entering port $2$, the output is of the form: \begin{eqnarray}
|12\rangle &\to & {1\over 4}\left( -|1\rangle  + |2\rangle + |3\rangle + |4\rangle \right) \\ &  &\qquad \cdot\left( +|1\rangle  - |2\rangle + |3\rangle + |4\rangle \right) \nonumber \\ &=& {1\over 4} \left( -|11\rangle + |12\rangle - |13\rangle - |14\rangle \right.  \\ & & \qquad +|21\rangle - |22\rangle + |23\rangle + |24\rangle  \nonumber  \\
& & \qquad +|31\rangle - |32\rangle + |33\rangle + |34\rangle \nonumber  \\ & & \qquad \left. +|41\rangle - |42\rangle + |43\rangle + |44\rangle\right)\nonumber   \\
&=& {1\over 4} \left( -|11\rangle +2|12\rangle -|22\rangle +|33\rangle +|44\rangle +2|34\rangle \right)\nonumber  \\
&=&  {1\over 4} \left( -\sqrt{2}|2,0,0,0\rangle +2|12\rangle -\sqrt{2}|0,2,0,0\rangle \right. \\ & & \qquad \left. +\sqrt{2}|0020\rangle +\sqrt{2}|0002\rangle +2|34\rangle \right) ,\nonumber
\end{eqnarray}  where, for example, $|2,0,0,0\rangle$ represents the Fock state with two-photons in port $1$, and zero photons in ports $2$, $3$, and $4$.
The extra factors of $\sqrt{2}$ in the last line come from writing the states with two photons in the \emph{same} port in terms of the two-particle Fock state in port $n$ with standard normalization; for example, \begin{equation}|11\rangle = a^\dagger a^\dagger |0\rangle = \sqrt{2}|2,0,0,0\rangle .\end{equation}

\begin{figure}
\centering
\includegraphics[totalheight=2.2in]{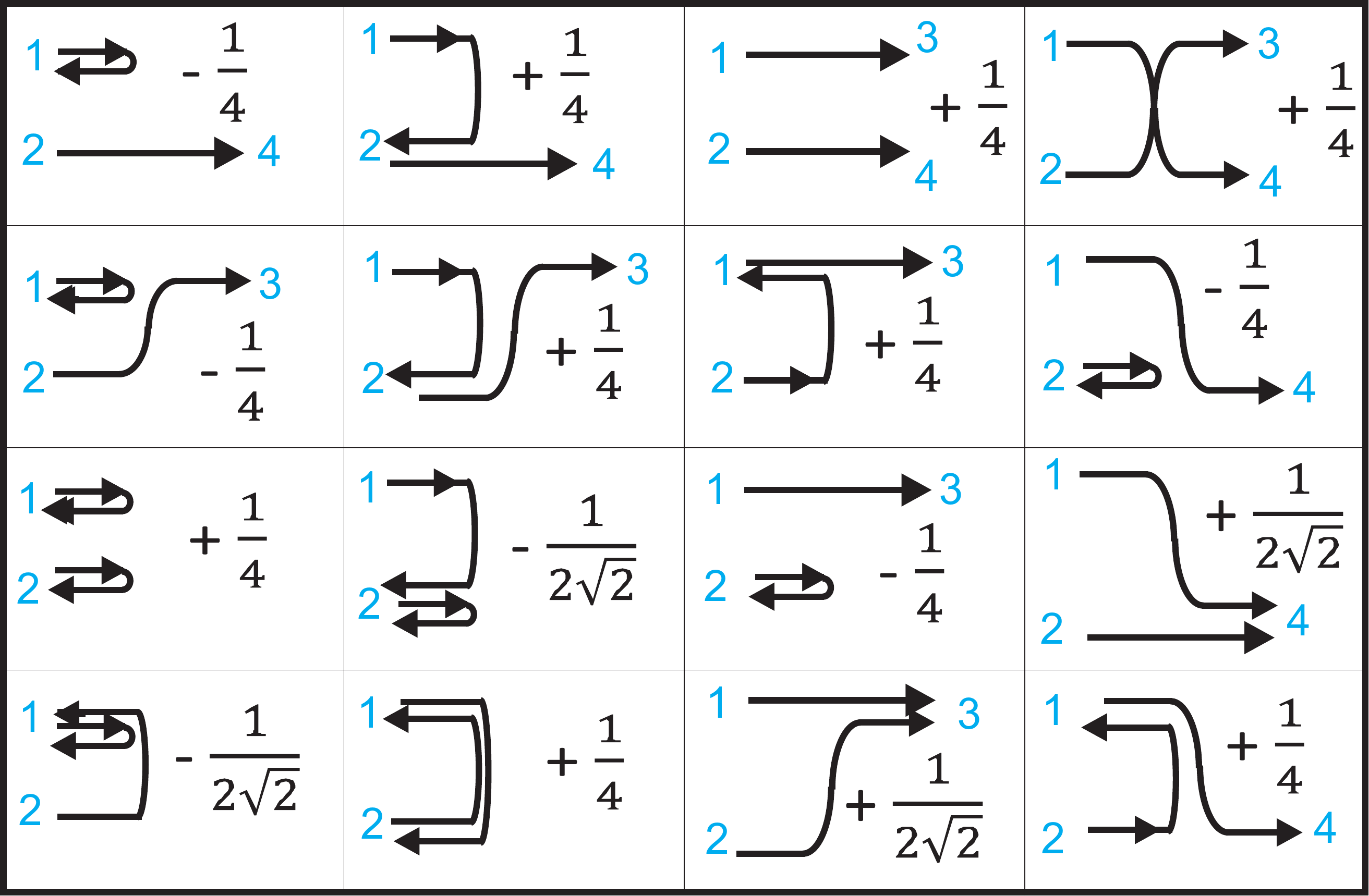}
\caption{Amplitudes for all the two-particle paths through a single four-port.}\label{amplitudefig}
\end{figure}

The transitions amplitudes from initial input state \begin{equation}|\psi_0\rangle =  |12\rangle \end{equation} to all possible output states are then easy to tabulate and are shown in the table of Fig. \ref{amplitudefig}.

\begin{figure}
\centering
\includegraphics[totalheight=.8in]{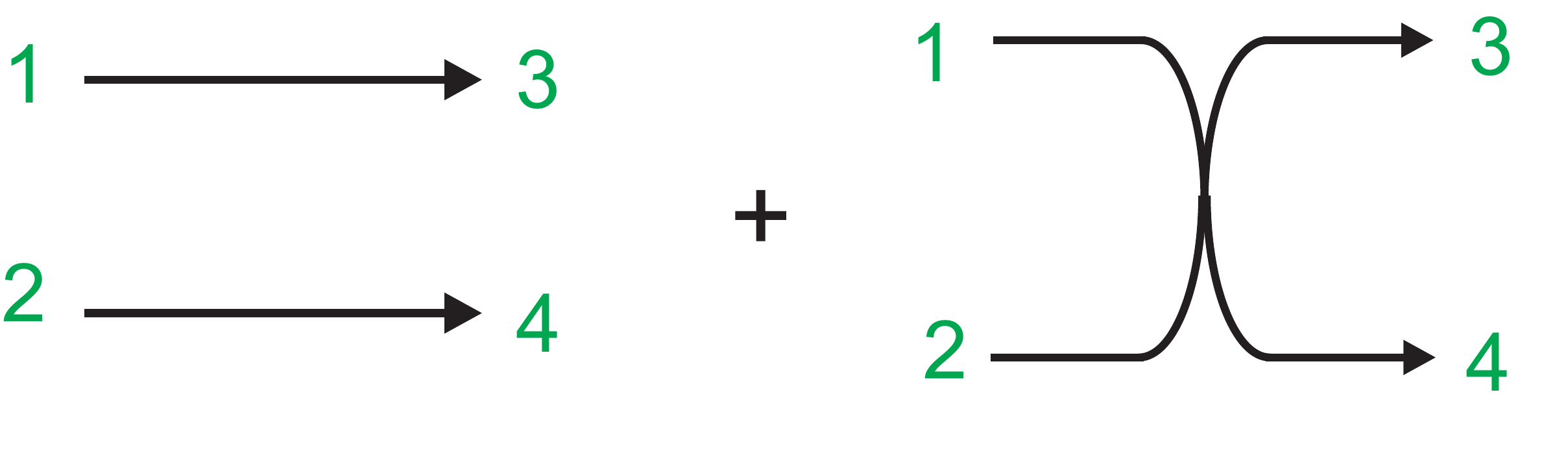}
\caption{Amplitudes of indistinguishable outcomes must be added; shown here the amplitudes for exit at ports $3$ and $4$ (the two cells in the top right corner of Fig. \ref{amplitudefig}), which add and then square to give the probability $P(|\psi_0\rangle \to |34\rangle ) $ in Table \ref{exitprobtable}.}\label{ampexfig}
\end{figure}

Exit probabilities can be found by adding the amplitudes for indistinguishable outcomes and squaring. For example, the probability for the photons to exit at ports $3$ and $4$ is given by the squared sum of two amplitudes (Fig. \ref{ampexfig}): $$P(|\psi_0\rangle \to |34\rangle ) ={1\over 2}\left( {1\over {2\sqrt{2}}}+ {1\over {2\sqrt{2}}}\right)^2 = {1\over 4}.$$
Table \ref{exitprobtable} lists the amplitudes for all possible output states of the Grover coin four-port, assuming input state $|\psi_0\rangle .$  %Directional labels are omitted from the states, but the exit directions are given in the last column.

The probabilities of both photons exiting left, both right, or of one in each direction are then found by simply adding the probabilities from the table,
\begin{eqnarray} P(\psi_0\to LL) &=& {1\over 8}+{1\over 4}+{1\over 8} ={1\over 2}\\
P(\psi_0\to RR) &=& {1\over 8}+{1\over 4}+{1\over 8} ={1\over 2}\\
P(\psi_0\to LR) &=& 0 ,
\end{eqnarray} as claimed in the main text (Eqs. \ref{eq1}-\ref{eq3}).

\begin{table}\label{probtable}
\centering % used for centering table
{\renewcommand{\arraystretch}{1.8}
\begin{tabular}{|c|c|c|}\hline
Transition & Probability & Exit Direction\\ \hline
\; $|\psi_0\rangle \to |11\rangle $ \; & \; $ P=\left( -{1\over {2\sqrt{2}}}\right)^2={1\over 8}$\;  &  LL\\  \hline
\; $|\psi_0\rangle \to |12\rangle $ \; & \; $ P=\left( {1\over 4}+{1\over 4}\right)^2={1\over 4}$\;   & LL\\ \hline
\; $|\psi_0\rangle \to   |13\rangle $ \; & \; $ P=\left( {1\over 4}-{1\over 4}\right)^2=0 $\;  & LR\\ \hline
\; $|\psi_0\rangle \to  |14\rangle $\;  & \; $ P=\left( {1\over 4}-{1\over 4}\right)^2=0 $\;  & LR\\ \hline
\; $|\psi_0\rangle \to  |22\rangle $ \; & \; $ P=\left( -{1\over {2\sqrt{2}}}\right)^2={1\over 8}$\;   & LL\\ \hline
\; $|\psi_0\rangle \to   |23\rangle $ \; & \; $ P=\left( {1\over 4}-{1\over 4}\right)^2=0 $\;  & LR\\ \hline
\; $|\psi_0\rangle \to   |24\rangle $ \; & \; $ P=\left( {1\over 4}-{1\over 4}\right)^2=0 $\;  & LR\\ \hline
\; $|\psi_0\rangle \to   |33\rangle $ \; & \; $ P=\left( +{1\over {2\sqrt{2}}}\right)^2={1\over 8} $\;  & RR\\ \hline
\; $|\psi_0\rangle \to  |34\rangle $ \; & \; $ P=\left( {1\over 4}+{1\over 4}\right)^2={1\over 4} $\;  &  RR\\  \hline
\; $|\psi_0\rangle \to |44\rangle $ \; & \; $ P=\left( +{1\over {2\sqrt{2}}}\right)^2={1\over 8} $\;  & RR \\ \hline
\end{tabular}}
\caption{Two-photon exit probabilities for input $|\psi_0\rangle$. } % title of Table
\label{exitprobtable} % is used to refer this table in the text
\end{table}

\section{Action of Grover Coin on Entangled States}

First, recall the standard maximally-entangled Bell states:

\begin{eqnarray}|\Psi^\pm \rangle &=& {1\over \sqrt{2}}\left( |1H\rangle |2V\rangle \pm  |1V\rangle |2H\rangle  \right) \\
|\Phi^\pm \rangle &=& {1\over \sqrt{2}}\left( |1H\rangle |2H\rangle \pm  |1V\rangle |2V\rangle  \right) ,
\end{eqnarray}
where, for example, $|jH\rangle$ represents a horizontally polarized photon in port $j$.

Then define an additional set of states:
\begin{eqnarray} |A_\pm^{ij} \rangle &=& {1\over \sqrt{2}} \left(|iH\rangle |jV\rangle \pm  |iV\rangle |jH\rangle \right) \\
|B_\pm^{ij} \rangle &=& {1\over \sqrt{2}} \left(|iH\rangle |iV\rangle \pm  |jV\rangle |jH\rangle \right) \\
|C_\pm^{ij} \rangle &=& {1\over \sqrt{2}} \left(|iH\rangle |jH\rangle \pm  |iV\rangle |jV\rangle \right)  .
\end{eqnarray}

These have the following properties:
\vskip 5pt

$\bullet$ If $i=j$:  $|A_+^{jj}\rangle =|B_+^{jj}\rangle$, and $|A_-^{jj}\rangle =|B_-^{jj}\rangle =0.$
\vskip 5pt

$\bullet$ For $i\ne j$, $|C_\pm^{12}\rangle = |\Phi^\pm\rangle $,  $|A_\pm^{12}\rangle = |\Psi^\pm\rangle $ are maximally entangled Bell states.
\vskip 5pt

$\bullet $ $ |C_\pm^{11}\rangle$ and $ |C_\pm^{11}\rangle$ can be seen as $N00N$ states with $N=2$ polarized photons.

We now examine how these states behave under the action of the Grover four-port, and see for which initial states two-photon clustering occurs.

\vskip 15pt

{\bf I.} The $A^{12}_+ $ state is already studied in Section \ref{entsection}. We know that at the first step it splits into a \emph{sum} of reflecting and transmitting states:
\begin{equation} |A^{12}_+\rangle \to |\psi_t\rangle +|\psi_r\rangle .\end{equation} The two photons in the reflecting and transmitting states then continue to cluster in all subsequent steps.

{\bf II.} In a similar manner, $B^{12}_+ $ splits into a \emph{difference} of reflecting and transmitting states:
\begin{equation} |B^{12}_+\rangle \to |\psi_t\rangle -|\psi_r\rangle ,\end{equation} again leading to photon clustering.

{\bf III.} The product state $A_+^{11}=\sqrt{2}|1H\rangle |1V\rangle $ transforms under the Grover coin at the first step as
\begin{eqnarray}|A_+^{11}\rangle &\to & {1\over {2\sqrt{2}}} \left[ |1H\rangle |1V\rangle +|2H\rangle |2V\rangle \right. \\
& & \qquad\quad +|3H\rangle |3V\rangle +|4H\rangle |4V\rangle \nonumber \\ & & \qquad\quad -\left( |1H\rangle |2V\rangle +|2H\rangle |1V\rangle\right)\nonumber \\
& & \qquad\quad -\left( |1H\rangle |3V\rangle +|3H\rangle |1V\rangle \right)\nonumber \\ & & \qquad\quad  - \left( |1H\rangle |4V\rangle +|4H\rangle |1V\rangle \right)\nonumber \\ & & \qquad\quad   +\left( |2H\rangle |3V\rangle +|3H\rangle |2V\rangle\right) \nonumber\\ & & \qquad\quad +\left( |2H\rangle |4V\rangle +|4H\rangle |2V\rangle \right)\nonumber \\
& & \qquad\quad \left. + \left( |3H\rangle |4V\rangle +|4H\rangle |3V\rangle \right) \right] \nonumber .\end{eqnarray} Since cross terms occur that mix left-moving (1 and 2) exit states with right-moving (3 and 4) states, no clustering occurs. $|A_+^{22}\rangle$ behaves similarly.

{\bf IV.} $A_-^{12} $ input leads to output at the first step of the form:
\begin{eqnarray}|A_-^{12}\rangle &\to &   {1\over {2\sqrt{2}}} \left[ \left( |1V\rangle |4H\rangle -|1H\rangle |4V\rangle  \right) \right. \\
& & \qquad + \left( |1V\rangle |3H\rangle -|1H\rangle |3V\rangle\right)  \nonumber\\ & & \qquad + \left( |2H\rangle |3V\rangle -|3H\rangle |2V\rangle\right) \nonumber \\
& & \qquad \left. + \left( |2H\rangle |4V\rangle -|2V\rangle |4H\rangle \right) \right]\nonumber  .\end{eqnarray} Not only does clustering not occur, but the two photons always go in \emph{opposite} directions at this first step. (This repulsion of course does not continue on subsequent steps, since the interference no longer occurs once the photons are separated.)

{\bf V.} The $C_{11}^\pm$ states transform at the first step as follows:
\begin{eqnarray}|C_\pm^{11}\rangle &\to & {1\over {4\sqrt{2}}}\left( |1H\rangle |1H\rangle +|2H\rangle |2H\rangle\right.  \\
& & \qquad\quad\left. +|3H\rangle |3H\rangle +|4H\rangle |4 H\rangle\right) \nonumber \\ & &  + {1\over {2\sqrt{2}}} \left[ \left( |2H\rangle |3H\rangle +|2H\rangle |4H\rangle  +|3H\rangle |4H\rangle\right)\right. \nonumber\\ & & \qquad \left.
-\left(  |1H\rangle |2H\rangle +|1H\rangle |3H\rangle  +|1H\rangle |4H\rangle \right) \right] \nonumber\\
& & \pm \mbox{ same with } H\to V.\nonumber
\end{eqnarray}
This does not cluster because of the crossed terms mixing 1 and 2 with 3 and 4 in the second line. $C_{22}^\pm$ behaves in a similar manner.

{\bf VI.} $C_\pm^{12}$ becomes a linear combination of all of the $C$ states under the action of the four-port:
\begin{eqnarray}|C_\pm^{12}\rangle &\to& {1\over 4} \left( -|C_\mp^{11}\rangle -|C_\mp^{22}\rangle +|C_\pm^{33}\rangle +|C_\pm^{44}\rangle\right) \nonumber \\
& & \qquad +{1\over 4}\left( |C_\pm^{12}\rangle +|C_\pm^{34}\rangle\right) .\end{eqnarray} In subsequent steps, this state then declusters because the $C_\pm^{jj}$ states in the first parentheses decluster, as shown in IV above.

So, summarizing, the  $A^{12}_+$ and $B^{12}_+$ states will always cluster, while the remainder of the $A$, $B$, and $C$ states will not.

\end{document}